\documentclass[11pt,aps,nofootinbib,preprintnumbers,citeautoscript]{revtex4}
\usepackage{amsmath,amssymb}
\usepackage[colorlinks=true,citecolor=red,linkcolor=blue]{hyperref}
\usepackage{breqn}
\usepackage{enumerate}
\usepackage{graphicx}
\usepackage{dcolumn}
\usepackage{bm}
\usepackage[mathlines]{lineno}
\usepackage[capitalize]{cleveref}
\usepackage{subfigure}
\usepackage{float}
\DeclareGraphicsExtensions{.pdf,.eps}

\begin{document}
\title{Milestone Developments in Quantum Information and No-Go Theorems.}

\author{Kapil K. Sharma$^{1}$, Vladimir P. Gerdt$^{2,3}$ and Pyotr V. Gerdt$^{2}$\\ \vspace{4mm}
\textit{$^{\{1\}}$ DY Patil International University, \\
Sect-29, Nigdi Pradhikaran, Akurdi, \\Pune, Maharashtra-411044, India.}\\
E-mail: $^{\{1\}}$\textit{iitbkapil@gmail.com}\\ \vspace{2mm}
\textit{$^{\{2\}}$Joint Institute for Nuclear Research, \\ 6 Joliot-Curie St, 141980 Dubna, Russia \\ 
E-mail: $^{\{2\}}$gerdt@jinr.ru, $^{\{2\}}$gapon1970@yandex.ru  \\
 \vspace{1mm} 
and  \vspace{1mm} \\
$^{\{3\}}$Peoples' Friendship University of Russia (RUDN University) \\
6 Miklukho-Maklaya St, 117198 Moscow, Russia} \\ 
}

\begin{abstract}
In this article we present milestone developments in the theory and applications of quantum information from historical perspectives. The domain of quantum information is very promising to develop quantum computer, quantum communication and varieties of other applications of quantum technologies. We also give the light on experimental manifestations of major theoretical developments. In addition, we present important  no-go theorems frequently used in quantum information along with ideas of their respective mathematical proofs.
\end{abstract}
\maketitle

\section{Introduction}\label{sec:introduction}

A breakthrough in the classical information theory has been done by C.Shannon in 1948 \cite{sh1}. This theory has its practical manifestations in communication systems, computing devices, gaming, imaging and with countless applications in real world.
The theory of quantum information, which generalizes Shannon's theory stems from the postulates of quantum mechanics and based on such fundamental ingredients of quantum mechanics as superposition and entanglement. Based on quantum information theory, the efforts to develop quantum computer are on the way by using several physical techniques. Many companies in the market are eagerly applying efforts to push this area towards commercialization and to develop quantum computer with different physical approaches like super conducting approach, ion trap system, ultra cold atoms, Majorana fermions, etc (see recent review~\cite{Bishnoi:2020} and its bibliography). A number of quantum computers with several dozens qubits have been created at the laboratory level, and in January 2019 IBM \url{(https://www.ibm.com/quantum-computing/)} announced the first commercial quantum computer. We may fire a natural question: can we store and process the information in these physical systems more efficiently than classical one? And what are the physical constraints responsible to execute quantum information~\cite{info1,info2}? So, investigating situations which are not possible is also an important paradigm. These impossible physical conditions are expressed by no-go theorems~\cite{nogo1,nogo2}. Before applying the fruitful efforts to develop any quantum application it is always good to keep in view the structure of no-go theorems, which is always helpful to tackle the feasibility and non-feasibility of physical situations. On the other hand, towards development of quantum computing, there are always challenges to manipulate and control the qubits and to protect them from decoherence. The phenomenon of decoherence is the killer of superposition in quantum systems and a serious restriction on the way to perfect quantum computation. Recent developments in quantum computation are very progressive and more rapid than the past historical developments. So, in this direction it is very important to understand the gradual milestone developments and track them what may be useful for further progress in the future. In the following sections, we present some major developments with theoretical aspects and touching the experimental issues as well. We managed the time periods of developments in two slots (1970-2000) and (2001-2020). 
\section{Duration (1970-2000)}
This period is most significant for the theoretical development of quantum information and known for producing the idea of reversible computation by C.Bennett~\cite{rv} and also by T.Toffoli~\cite{tofo} who invented the first reversible n-bit quantum gates which are heavily used in the circuit model of quantum computation. Another milestone development is the foundation of Holevo bound. A.Holevo has established the upper bound for amount of information that can be contained in a quantum system~\cite{hv}.  Then one of the first attempts to create the quantum information theory was made by R.Ingarden, a Polish mathematical physicist, who published a seminal paper entitled ``Quantum information theory" in 1976~\cite{qi1}. This work generalizes Shannon's information theory~\cite{sh1} for quantum mechanics of open systems. With the progress of quantum information theory, the idea towards quantum computing was proposed by Yu.Manin~\cite{qc} in 1980. Then R.Feynman argued~\cite{Feynman:1982} that for simulation of quantum systems one should use computational devices based on quantum physics, since such simulation tasks are very hard for classical computers. In 1982, a major result of no-cloning in quantum physics was discovered by W.Wootters and W.Zurek~\cite{nc1} and independently by D. Dieks~\cite{nc2}. The no-cloning theorem states that it is not possible to clone arbitrary quantum states. This theorem became the milestone for quantum information. We are inclined to present this theorem with its proof in Sect. 4.1. With this majour development, P.Benioff proposed a  first theoretical model for quantum computation based on quantum Hamiltonian~\cite{qm}. He made the first attempt to quantize the Turing machine and the framework of quantum Turing machine has taken place. The concept of entanglement has already taken birth during 1935 and 1936 with the debate of  A.Einstein and  E.Schrödinger \cite{entan1,entan2}. The advantage of entanglement and no-cloning theorem together captured the discovery of the first protocols for quantum cryptography (quantum key distribution) done by Ch.Bennet and G.Brassard in 1984~\cite{BB:84} and A.Ekert in 1991~\cite{AkEkert1991}. The development of quantum cryptography opened the new field of secure quantum communication, which is very promising to this date.

The last decade of the XX century extensively contributed to the development of entanglement - based quantum algorithms. In 1992, D.Deutsch and R.Jozsa proposed a deterministic quantum algorithm to test weather a function is balanced or constant by using black box model in quantum computation~\cite{drj}. With continuation of this work, the first milestone quantum algorithm was formulated by P.Shor at Bell Labs (New Jersey) in 1994 and published in 1997~\cite{fact}. The algorithm allows a quantum computer to factor an integer number very fast and runs in polynomial time. This algorithm can easily break the public-key cryptographic schemes as RSA scheme~\cite{rsa}. Meanwhile, as the development on quantum algorithms, P.Shor and A.Steane proposed the schemes for quantum error corrections in 1995~\cite{e1,e2}. Quantum error corrections protocols are used to protect quantum information from decoherence and essentially needed in ``noisy'' quantum hardware. After design of Shor's algorithm, L.Grover in 1996~\cite{dd} invented the quantum search algorithm in an unsorted database. That algorithm is the fastest one and has become a landmark in quantum computation. Here we mention that the period of (1990-1997) has been recognized as the golden period for theoretical as well as experimental developments in quantum computation. Beside the quantum algorithms development, there also was discovered  important protocol of quantum information processing  called quantum teleportation.  It was  proposed by C.Bennett et al. in 1993~\cite{tele1} and experimentally verified in 1997~\cite{tele2}. During 1997 the scientific community was strongly focused on experimental manifestations of quantum information around the world. The first experimental approach to realize the quantum gates by using nuclear magnetic resonance (NMR) technique was performed by N.Gershenfeld and I.Chuang in 1997~\cite{q2}. NMR technique came out as a useful resource to produce fruitful experimental manifestations of quantum computation. In 1998, the first execution of  Deutsch-Jozsa algorithm was performed by using just NMR technique. This has been done by J.Jones and M.Mosca at Oxford University and shortly later by I.Chuang at IBM's Almaden Research Center together with co-workers at Stanford University and MIT~\cite{dj}. In the same year Grover's algorithm was also experimentally verified with NMR quantum computation~\cite{mi}. This experimental development encouraged the further investigations. Beside the theoretical and experimental manifestations, it was also major interest to look into some physical situations which are not feasible like the non-cloning theorem. Towards this direction, there is one important quantum no-deletion theorem proved by A.Pati and S.Braunstein~\cite{ng}, which states that, given two copies of arbitrary qubit states, one cannot delete one of them if it is unknown. This theorem has its own important implications in quantum information~\cite{ng}. We are inclined to discuss this theorem in Sect 4.4.
\section{Duration (2001-2020)}
This period is well known for the role of quantum optics in quantum information, in parallel with another major developments towards the implementation of quantum networks. In 2001, the first experimental execution of Shor's algorithm at IBM's Almaden Research Center and Stanford University was implemented by using NMR technique~\cite{ed}. The number $15$ was factored by using $10^{18}$ identical molecules. In the same year, the scenario of optical quantum computing has been started. E.Knill, R.Laflamme and G.Milburn showed that optical quantum computing is possible with beam splitters, phase shifters, single photon sources and photo-detectors~\cite{rs}. They also have shown that quantum teleportation can be performed with beam splitters by using photonic qubits. Their contribution opened the avenues of usage of optics in quantum information. The role of optics is very promising nowadays to establish long distance quantum communication. The implementation of quantum gates with optics is an essential requirement to perform quantum computation. In this direction,  quantum Controlled-Not gates using linear optical elements has been developed by T.Pittman and collaborators at  Applied Physics Laboratory,  Johns Hopkins University in 2003~\cite{op1}. The similar results have been produced independently by J.O'Brien and collaborators at the University of Queensland~\cite{op2}. Quantum optics not only had its applications in quantum cryptography but DARPA Quantum network also became operational by using optical fibers supporting the transmission of entangled photons~\cite{dar}. Quantum networks use the protocol called quantum repeater for long distance quantum communication to overcome decoherence. These quantum repeaters transmit the quantum states to receiver with the help of quantum memories. The recognizable framework of quantum optics with atom-photon interaction proved to be a successful  framework and assisted to develop quantum memories~\cite{qm1}, which are essential to establish quantum Internet~\cite{qint}. In 2005, at Harvard University and Georgia Institute of Technology researchers succeeded in transferring quantum information between ``quantum memories" from atoms to photons and back again~\cite{in}. Along with the advancement of quantum networks, the concept of distributed  quantum computing has taken place and a protocol called quantum telecloning was proposed by M. Murao et al. in 1999~\cite{tele}. This is the protocol in which the optical clones of an unknown quantum state are created and distributed over distant parties. Of course, the quantum no-cloning theorem implies that these copies cannot be perfect. S.Braunstein at the University of York together with his colleagues from the University of Tokyo and the Japan Science and Technology agency gave the first experimental demonstration of quantum telecloning in 2006~\cite{qtc}. Quantum networks and quantum repeaters attracted much attention of quantum community, and along this line of research the concept of entanglement swapping~\cite{qrepet} was developed by S.Pirandola et al. in 2006 which has its important application in quantum repeaters. Beside the developments on quantum memoriy by using the optical techniques, there was also interest to develop such memory by using the condensed matter approach. It was done in 2007 by using the Bose-Einstein condensation \cite{bed}. Till 2007, the experimental manifestation of two-qubit entanglement has been successfully performed, but the entanglement in hybrid systems also attracted attention of quantum community. Much progress was done in 2008 to perform photonic qubit-qutrit entanglement~\cite{gr}. In the direction of implementation of quantum networks and towards the reality of quantum Internet, the logic gates have been implemented in optical fibers by P.Kumar, which became the foundation of quantum networks~\cite{qn}.

The continuity of past developments in quantum information and its experimental manifestations were maintained in this era with two major center of interest: how to develop efficient quantum processors and how to increase the coherence time in quantum systems. On the other hand, few past records also broken in this era. As a continuation, in 2011, the von Neumann`s architecture was employed in quantum computing with superconducting approach~\cite{von}. This work contributes in developing quantum central processing unit that exchanges  the data with a quantum random-access memory integrated on a chip. There was a breakthrough in 2014, as scientists transferred data by quantum teleportation over the distance of 10 feet with zero percent error rate, this was a vital step towards a feasible quantum internet~\cite{tp}. In the same year, N.Dattani and N.Bryans broken the record for factoring the unbeatably largest number 56153 using NMR with 4 qubits only on a quantum device which outperformed the record established in 2012 with factoring the number 143~\cite{fc1}. After a long journey for development in quantum computation, there are still many theoretical and experimental open problems inherited in the essence of quantum information. One of the major issues is controlling entanglement and its manipulation in many-body quantum systems with its protection from decoherence. There have been successful  efforts to increase the coherence time in 2015 to six hours in nuclear spins~\cite{dc1}. With the advancement of quantum processors, there was breakthrough in 2017 by D-wave. This company developed commercially available quantum annealing based quantum processor, which is fully functional now and have been used for varieties of optimization problems~\cite{qa} with applications in quantum machine learning. The recent progress in quantum computing hardware is described in~\cite{Bishnoi:2020}. With the connection of improving coherence time and deeper theoretical investigations on entanglement, here we mention that entanglement is a fragile phenomenon and very sensitive to quantum measurements and environmental interactions. It may die for a finite time in a quantum system and alive again as time advances. This phenomenon is called entanglement sudden death (ESD) which was observed earlier by Yu-Eberly~\cite{yu1,yu2} and investigated in \cite{ed1}-\cite{ed7}. The phenomenon of ESD is a threat to quantum applications, so overcoming from it is again an issue and needs fruitful solutions. During the period 2011-2020, there have been vast  research on entanglement and related aspects such as  distillable entanglement and bound entanglement in quantum information theory initiated in~\cite{db}. Quantum community has investigated various mathematical tools of entanglement detections and quantification, distillable protocols, monogamy of entanglement. However, these aspects are lacking for higher dimensional quantum systems. The efforts of quantum community is always to search the  quantum systems which can sustain long coherence time, which is an important topic of research. For more recent advances in quantum information and communication we refer to the review~\cite{RoadMap:2018}. We single practical achievements in quantum cryptography~\cite{Diamanti:2016,Bedington:2017}.
\section{No-go theorems}
In this section we consider important no-go theorems in theory of  quantum information~\cite{Pathak}. A no-go theorem implies the impossibility of a particular physical situation~\cite{nogo1,nogo2}. These theorems have the major impact on the experimental development of quantum information. All these theorems are developed by taking the linear property of quantum mechanics. Here in the following subsections we discuss important no-go theorems with their corresponding proofs.
\subsection{No-cloning theorem}
The no-cloning theorem states that one can not create an identical copy of an arbitrary pure quantum state. No-cloning theorem is provided by L.Park in 1970~\cite{Park:1970}, then further re-investigated in 1982 by W.Wootters and W.Zurek~\cite{nc1} and separately by D.Dieks~\cite{nc2}. The theorem of quantum cloning is easy to prove. Here we give a proof of this theorem based on unitarity of the underlying operation.
Consider two pure states as $|\psi\rangle$, $|\phi\rangle$ and a blank state $|b\rangle$. Mixing each pure state with blank state and perform the unitary operation which has the goal to copy the pure state into a blank state. So we get,
\begin{equation}\label{nc22}
U(|\psi\rangle\otimes|b\rangle)=|\psi\rangle\otimes|\psi\rangle\,,\qquad U(|\phi\rangle\otimes|b\rangle)=|\phi\rangle\otimes|\phi\rangle\,.
\end{equation}
Taking the complex conjugate of both the sides of both the above equations
we obtain
\begin{equation}
(\langle\psi|\otimes\langle b|)U^{\dagger}=\langle\psi|\otimes\langle\psi|\,.\label{nc3}
\end{equation}
By multiplying the left and right sides of Eq.\,\eqref{nc22} and Eq.\,\eqref{nc3} we find
\[
(\langle\psi|\otimes\langle b|)U^{\dagger}U(|\phi\rangle\otimes|b\rangle)=(\langle\psi|\otimes\langle\psi|)
(|\phi\rangle\otimes|\phi\rangle)\,.
\]
We know that $U^{\dagger}U=I$ and therefore
\[
\langle\psi|\phi\rangle=\langle\psi|\phi\rangle^{2}\,.
\]
The equation is conflicting and it is true only if $\langle\psi|\phi\rangle=0$,
or $|\psi\rangle=|\phi\rangle$. These conditions reveal that there is no unitary operation which can be used to clone an arbitrary
quantum state. 
\subsection{No-broadcast theorem}
This theorem generalizes the no-cloning theorem to mixed states. The first attempt to prove that non-commuting mixed states can no be broadcast was done by H. Barnum et al. in~\cite{br1}. Further extensions are done by many authors. To look into broad view of no-broadcasting and different paradigms, we refer to  \cite{br1}-\cite{br5}. Before formulation of the theorem we explain what is  meant by broadcasting (cf.~\cite{Pathak}). Let we have a mixed state $\rho$ in $H$ and a map $E$ that maps $\rho$ to a state $\rho_{AB}$ on $H_A\otimes H_B$. Then $E$ broadcasts $\rho$ if
\begin{equation}\label{cond:broadcast}
     \text{Tr}_A(\rho_{AB})=\text{Tr}_B(\rho_{AB})=\rho\,,
\end{equation}
where $\text{Tr}_A$ and $\text{Tr}_B$ denote partial trace over the subsystem $A$ and subsystem $B$, respectively. Note that the final state $\rho_{AB}$ is not
necessarily a product state.

The no-broadcast theorem states that
no such map $E$ exists for an arbitrary quantum state, i.e., arbitrary quantum states cannot be broadcasted. Alternatively, given an
arbitrary quantum state $\rho$, it is impossible to create a state $\rho_{AB}$ such
that the equality~\eqref{cond:broadcast} holds. Moreover, the two mixed states $\rho_1,\rho_2$ in $H$ can satisfy~\eqref{cond:broadcast} if and only if they commute~\cite{br1}.

We outline here the main idea of the proof given in~\cite{br4} that is the most general one among known proofs and relies on fundamental principles of information theory, especially on entropy-based arguments. This proof allows to establish a link of the quantum theorem to its classical analogue~\cite{DPP:2002}.

The relative entropy $S(\rho_1|\rho_2)=\text{Tr}[\rho_1(\log\rho_1-\log\rho_2)]$ of states $\rho_1$ and $\rho_2$ is conserved under unitary time evolution
\begin{equation}\label{cons:entropy}
S(\rho_1(t)|\rho_2(t))=S(\rho_1(0)\mid \rho_2(0))
\end{equation}
and it is monotone, i.e.,
\begin{equation}\label{monotonicy}
S(\rho_{1,AB}|\rho_{2,AB})\geq S(\rho_{1,B}|\rho_{2,B})\,,
\end{equation}
where $\rho_{1,AB}$, $\rho_{2,AB}$ are density operators of the composite system $AB$ and $\rho_{1,B}$, $\rho_{2,B}$ are the corresponding density operators of the subsystem $B$. From\,Eqs.\,\eqref{cons:entropy} it follows
\[
  S(\rho_1|\rho_2)=S(\rho_{1,AB}|\rho_{2,AB})\,,
\]
whereas for non-commuting $\rho_1$ and $\rho_2$.  On the other hand, Eq.\,\eqref{monotonicy} implies the strict inequality
\[
  S(\rho_1|\rho_2) < S(\rho_{1,AB}|\rho_{2,AB})\,,
\]
a contradiction.
\subsection{No-deletion theorem}
The no-deletion theorem states~\cite{ng} that, given two copies of an arbitrary pure quantum state, it is impossible to delete one of them by a unitary interaction.

Let we define the quantum deleting machine as follows
\begin{equation}\label{eq:deletion}
(\,\exists\, U,A,A^\prime)\ \left(\,\forall\, |\psi\rangle\,\right)\  \big[\,U|\psi\rangle|\psi\rangle|A\rangle=|\psi\rangle|0\rangle |A^\prime \rangle \,\big]
\end{equation}
On the left-hand side of the equation, $U$ a the unitary operator acting on the composited state $|\psi\rangle|\psi\rangle|A\rangle\in H\otimes H\otimes H_A$ where $|A\rangle$ is an ancilla state, which is independent of $|\psi \rangle$ . In the right-hand side of equation~\eqref{eq:deletion} the state $|0\rangle$ is a blank one of the same dimension as $|\psi\rangle$ signifying deletion of the last state, and $|A^\prime \rangle$ is the final state of the ancilla. If the equation in~\eqref{eq:deletion} holds, then
\[
  |\psi\rangle|\psi\rangle|A\rangle=U^{-1}|\psi\rangle|0\rangle |A^\prime \rangle\,.
\]
But this stands in contradiction to the no-cloning theorem. 
\subsection{No-teleportation theorem}
In quantum information, the no-teleportation theorem\,\cite{nt1} states that neither an arbitrary quantum state can be converted into a sequence of classical bits nor the classical bits can create original quantum state. This theorem is the consequence of no-cloning theorem. If an arbitrary quantum states allow producing sequence of classical bits, then as we know the classical bits can always be copied and hence the quantum state also can be copied, which violate the no cloning theorem. So the conversion of an arbitrary quantum states in sequence of classical bits is not possible.

The similarity of two states is defined as follows. Two quantum states $\rho_{1}$ and $\rho_{2}$ are identical if the measurement results of any physical observable have the same expectation value for $\rho_{1}$ and $\rho_{2}$. Let prepare an arbitrary mixed quantum state $\rho_{input}$, then perform the measurement on the state and obtain the classical measurement results. Now by using these classical measurement results the original quantum states is recovered as $\rho_{output}$. Both the input and output states are not equal
\begin{equation}\label{r}
\rho_{input}\neq \rho_{output}\,.
\end{equation}
This result holds irrespective to the state preparation process and measurement  outcome. Hence Eq.\,\eqref{r} proves that one can not convert an arbitrary quantum states into a sequence of classical bits. 

This theorem does not have any relation to teleportation of a quantum state based on the phenomenon of quantum entanglement as a means of transmission (see, for example, textbooks~\cite{entan2,Pathak}).
\subsection{No-communication theorem}
No-communication theorem\,\cite{nct1,ncc2} is also known as the no-signaling principle.
This theorem essentially states that is not possible to transmit classical bits of information by means of carefully prepared mixed or pure states, whether entangled or not.
Therefore, the theorem disallows any communication, not just superluminal (i.e., faster than the speed of light in vacuum), by means of shared quantum states.

Let Alice and Bob share a quantum state. Our goal is to show that the measurement action performed at the end of Alice is not detectable by Bob at his end, and this is true in either case, weather the composite state of Alica and Bob is separable or entangled. We first consider the case when the composite state is separable.
Denote this state by $\rho$, and let Alice perform a measurement on her end. Such measurement can be modeled by Kraus operators, which may not commute. Denote the Kraus operator(s) at the end of Alice by $A_{m}$. Now the probability of measurement outcome $x$ can be written as
\begin{eqnarray*}
p_{x}=\sum_{m} Tr\big(A_{x,m}\, \rho\, A^{\dagger}_{x,m}\big)=Tr[\,\rho\, V_{x}\,]\,,\quad V_{x}=\sum_{m}A^{\dagger}_{x,m} A_{x,m}\,, \quad \sum_{x}V_{x}=1\,.
\end{eqnarray*}
Now denote the Kraus operator(s) at the end of Bob by $B_{n}$. The probability of measurement outcome $y$ at the end of Bob, irrespective to what Alice has found, is given by
\begin{equation}\label{ncc1}
p_{y}=\sum_{x}Tr\big(\sum_{m,n}B_{y,n}A_{x,m}\,\rho\, A^{\dagger}_{x,m}B^{\dagger}_{y,n}\big)\,.
\end{equation}
The order of measurements on the separable composite system does not matter, so the following commutation relation $[A_{x,m},B_{y,n}]=0$ holds, and Eq.\,\eqref{ncc1} can be rewritten as
\begin{equation*}
p_{y}=\sum_{x}Tr\big(\sum_{mn}A_{x,m}B_{y,n}\,\rho\, B^{\dagger}_{y,n} A^{\dagger}_{x,m}\,\Big)\,.
\end{equation*}
Using the cyclic property of trace operation and expanding the summation we obtain
\begin{equation*}
p_{y}=Tr\big(\sum_{n}B_{y,n}\,\rho\, B^{\dagger}_{y,n}\big)\,.
\end{equation*}
In this equation all the operators of Alice disappear, so Bob is not able to detect which Alice's measurements has been performed at her end. Hence, the statistics of measurements at the end of Bob has not been effected by Alice.

Suppose now that the composite state is entangled. Assume for simplicity that this state is the two-particle spin-singlet one
$
|\psi\rangle=\frac{1}{\sqrt{2}}(|01\rangle+|10\rangle)\,,
$
where $|0\rangle$ is the spin down state and $|1\rangle$ is the spin up state. Let Alice and Bob perform the measurement of the particles at their disposal by using the detectors $D^{A}$ and $D^{B}$, respectively. Following the Bell's experiment (see, for example, \cite{Benenti}, Sect.\,2.5.), assume that the detectors are initially oriented along the z axis and rotated independently at the end of Alice and Bob in such a way that the difference between the angles of detectors becomes $\theta$. Then it yields the conditional probabilities of measurements outcome (cf.\,\cite{Benenti},\,Eq.\,(2.182)):
\begin{eqnarray}
&&\{A(0), B(0)\}, \ P_{00}=\frac{1}{2}\sin^{2}\Big(\frac{\theta}{2}\Big)\,, \qquad
 \{A(0), B(1)\}, \ P_{01}=\frac{1}{2}\cos^{2}\Big(\frac{\theta}{2}\Big)\,,\\
&&\{A(1), B(0)\}, \ P_{10}=\frac{1}{2}\cos^{2}\Big(\frac{\theta}{2}\Big)\,, \qquad
 \{A(1), B(1)\}, \ P_{11}=\frac{1}{2}\sin^{2}\Big(\frac{\theta}{2}\Big)\,,
\end{eqnarray}
with the normalization condition
$
P_{00}+P_{01}+P_{10}+P_{11}=1\,.
$
Calculating the probabilities of the measurement outcome for spin up $(|1\rangle)$ and spin down $(|0\rangle)$ at the Alice end gives
$
P^{A}_{1}=P_{11}+P_{10}=\frac{1}{2}\,,\quad P^{A}_{0}=P_{01}+P_{11}=\frac{1}{2}\,.
$
Similarly, for the probabilities at the Bob end, we obtain  $P^{B}_{1}=P^{B}_{0}=\frac{1}{2}$. Thus, we the probabilities of measurement outcomes are totally independent on the angle $\theta$, and the actions performed by measurements at either end of Alice or Bob are not detected at another end and vice versa. This is the essence of no-communication theorem. 
\subsection{No-hiding theorem}
No-hiding theorem\,\cite{nhd1} is an important theorem in quantum information, which indicates the conservation of information. The idea of no-hiding theorem come out from the correlation of quantum information processing with the classical one in Vernam's cipher used in cryptography. This one-time pad cipher is based on addition of a random (secret) key to the original information to be transmitted. C.Shanon in\,\cite{{sh1}} proved that the original information neither reside in encoded message nor in the key, so where the information is gone? In the one-time pad cipher this information is hidden in the correlations with the key. One can think of the same scenario in quantum mechanical sense. The teleportation can be assumed as a quantum analogue of one-time pad cipher. In teleportation, there are two parties Alice and Bob both share an entangled state. Alice applies few unitary operations at her end and sends the measurement results to Bob. Bob applies the corresponding measurements and recovers the relevant information from Alice. In this whole process, the decoherence is not considered. If one considers the decoherence in teleportation process, then one has to take into account interaction with environment. If a quantum system interacts with an environment in the form of decoherence, then it destroys the original information. So a natural question arises: where the lost information from the original system has gone? In the quantum mechanical case it does not reside in correlations. This idea leads to the "No-hiding theorem". The theorem states that the original information resides in the subspace of the environmental Hilbert space and not in the part of correlation of the system and environment.

From a subspace $I$ we chose an arbitrary input mixed quantum state $\rho_{I}$ and encode it into a larger Hilbert space. With respect to a hiding process, there exists an output mixed state $\sigma_{O}$ of a subspace $O$ of the encoded Hilbert space. The remainder of the last space is regarded as an ancilla space $A$. The hiding process is characterized by the following mapping $\rho_{I}\longmapsto \sigma_{O}, \ (\sigma \ \text{is fixed}\ \forall \rho)$.

To be physical, the hiding process must be linear and unitary. Because of linearity, it is sufficient to  consider a pure input state $\rho_{I}=|\psi\rangle_{I\,I} \langle \psi |$. In this case the hiding process can be established as the Schmidt decomposition of the final state
\begin{equation}\label{map:hiding}
|\psi\rangle_I\mapsto\sum_{i=1}^{N}\sqrt{p_{i}} |i\rangle_O\otimes |A_i(\psi)\rangle_A.
\end{equation}
Here $p_{i}$ are nonzero eigenvalues of state $\sigma_{O}$ and $\{|i\rangle\mid 1\leq i\leq N\}$ are their  eigenvectors.
The sets $\{|i\rangle\}$ and $\{|A_{n}\rangle\}$ are orthonormal. By linearity,
\[
 |A_i(\mu |\psi\rangle+\nu |\psi_\perp\rangle)\rangle=\mu|A_i(\psi)\rangle+\nu|A_i(\psi_\perp)\rangle \,,
\]
where $|\psi_\perp\rangle$ denotes a state orthogonal to $|\psi\rangle$. Hence, the inner product of two such ancilla states gives
\[
  \mu^{*}\nu \langle A_m(\psi)|A_n(\psi_\perp)\rangle +\mu \nu^{*}\langle A_m(\psi_\perp)|A_n(\psi)\rangle=0\,.
\]
For arbitrary complex numbers of $\mu$ and $\nu$, all cross-terms must vanish. Thus, an orthonomal basis $\{|\psi_j\rangle\}$ defines an orthonomal set of states $|A_{n,j}\rangle=|A_n(\psi_j)\rangle$ spanning a Hilbert space that completely describes the reduced state of the ancilla. Hence, $|A_{n,j}\rangle=|q_n\rangle \otimes |\psi_j\rangle \oplus 0$, where $\{|q_n\rangle\}$ is an orthonomal set and $\oplus 0$ means that we pad unused dimensions of the ancilla space by zero vectors. Thereby, Eq.\,\eqref{map:hiding} takes the form
\[
|\psi_{I}\rangle=\sum_{i=1}^{n}\sqrt{p_{i}} |i\rangle_{O}\otimes (|q_{n}\rangle\otimes |\psi\rangle\oplus 0)_{A}\,.
\]
Since we may swap the state $|\psi\rangle$ with any other state in the ancilla using purely ancilla-local operations, we conclude that any information about $|\psi\rangle$ that is encoded globally is in fact encoded entirely within the ancilla. Neither the information about $|\psi\rangle$ is encoded in system-ancilla correlations nor in the system-system correlations. It is significant that this theorem has been verified experimentally~\cite{exper:no-hiding}.
\subsection{No-programming theorem}
This theorem states~\cite{NielsenChuang:1997} that no deterministic universal quantum processor, i.e. a processor which can be programmed to perform any unitary operation, can be realized.
Let distinct (up to a global phase) unitary operators $U_1,\ldots,U_N$ be implemented by some programmable quantum gate array. If so, the program register is at least $N-$dimensional, i.e., contains at least $\log_2N$ qubits. It is straightforward to show that the corresponding programs $|\mathcal{P}_1\rangle, \ldots,|\mathcal{P}_N\rangle$ are mutually orthogonal.\\[-0.5cm]
\begin{figure}
\begin{center}
\includegraphics[width=0.6\textwidth]{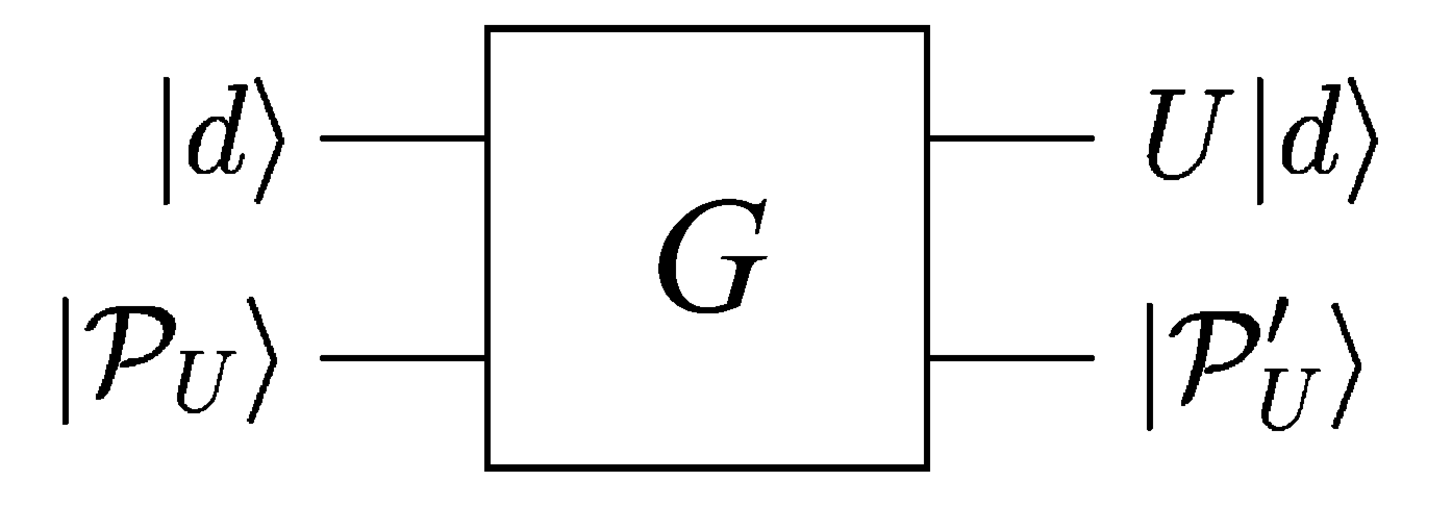} \\[0mm]
\end{center}
\vspace*{-8mm}
\caption{Schematic circuit for a programmable quantum
processor which implements the unitary operation $U$ determined
by the quantum program $|{\mathcal{P}}_U\rangle$.
\label{fig1}}
\end{figure}
Let $\mathcal{P}$ and $\mathcal{Q}$ be programs implementing unitary operators $U_p$ and $U_q$, respectively (see Fig.\,\ref{fig1}). Then for arbitrary data $|d\rangle$ we have equalities
\begin{equation}\label{programs}
   G\big(|d\rangle\otimes |{\mathcal{P}}\rangle\big) = (U_p|d\rangle)\otimes |{\mathcal{P}}^\prime\rangle\,,\quad  G\big(|d\rangle\otimes |{\mathcal{Q}}\rangle\big) = (U_q|d\rangle)\otimes |{\mathcal{Q}}^\prime\rangle\,.
\end{equation}
Taking the inner product of the equations in \eqref{programs} we obtain
\[
   \langle {\mathcal{Q}}|{\mathcal{P}}\rangle = \langle{\mathcal{Q}}^\prime|{\mathcal{P}}^\prime\rangle\langle d|U^\dag_qU_p|d\rangle\,.
\]
Assume that $\langle{\mathcal{Q}}^\prime|{\mathcal{P}}^\prime\rangle\neq 0$. Then the left-hand side does not depend on $|d\rangle$, and hence $U^\dag_qU_p=\gamma I$. It follows that if $\langle{\mathcal{Q}}^\prime|{\mathcal{P}}^\prime\rangle\neq 0$, then $U_p$ and $U_q$ are the same up to a global phase. This contradicts to our assumption on distinction of $U_p$ and $U_q$. Thus, $\langle{\mathcal{Q}}^\prime|{\mathcal{P}}^\prime\rangle = 0$ what implies $\langle {\mathcal{Q}}|{\mathcal{P}}\rangle=0$. This result demonstrate that no {\em deterministic} universal quantum processor exists. 

Although the deterministic universal quantum array cannot be realized, it is still possible to conceive it in an approximate (probabilistic) fashion~~\cite{NielsenChuang:1997}. In~\cite{Kubicki:2019} some bounds for the minimal resources necessary for this aim are given.
\section{Conclusions}
In this article, we discussed milestone developments in quantum information.
We captured the experimental manifestations of theoretical developments as well. In addition, we discussed physical situations, which are not possible in no-go theorems with their mathematical proof. These theorems have important consequences for quantum information processing and for quantum communication. Covering the broad aspects of milestone developments in this article may be useful for  quantum information community.
\section{Acknowledgements}
The publication has been prepared with the support of the ``RUDN University Program 5-100".

\end{document}